\documentclass[reprint,aps,amsmath,amssymb,superscriptaddress,prb]{revtex4-2}
\usepackage{graphicx}
\usepackage{dcolumn}
\usepackage{bm}
\usepackage{color}
\usepackage{amsmath}
\usepackage{booktabs} 
\usepackage{array} 
\usepackage{makecell}
\usepackage{colortbl}
\usepackage{xcolor} 
\usepackage[justification=centering,singlelinecheck=false]{caption}
\usepackage{braket}
\usepackage{subcaption}
\usepackage{caption}
\begin{document}
\title{Quantum-Electrodynamical Time-Dependent Density Functional
Theory Description of Molecules Interacting with Light}
\author{Yetmgeta Aklilu}
\affiliation{Department of Physics and Astronomy, Vanderbilt
University, Nashville, Tennessee, 37235, USA}
\author{Tiany Yang}
\affiliation{Department of Physics and Astronomy, Vanderbilt
University, Nashville, Tennessee, 37235, USA}
\author{Cody Covington}
\affiliation{Department of Chemistry, Austin Peay State University,
Clarksville, USA}

\author{K\'alm\'an Varga}
\email{kalman.varga@vanderbilt.edu}
\affiliation{Department of Physics and Astronomy, Vanderbilt
University, Nashville, Tennessee, 37235, USA}

\begin{abstract}
We study light-mediated interactions between spatially separated
molecules using real-time quantum electrodynamical time-dependent
density functional theory based on the Pauli–Fierz Hamiltonian. An
ultrashort delta-kick excitation selectively perturbs a single
molecule, while a second, distant molecule remains initially
unexcited. In free space, the excitation stays localized and no
response is observed in the second molecule. In contrast, when both
molecules are coupled to the same cavity mode, the initial excitation
induces coherent dynamics in the distant molecule through the shared
quantized electromagnetic field.
\end{abstract}

\maketitle

\section{Introduction}

Cavity quantum electrodynamics (cavity QED) explores how quantized
electromagnetic fields inside optical resonators reshape atomic and
molecular behavior. When molecules are placed between mirrors, the
vacuum field enhances or suppresses spontaneous emission, modifies
energy landscapes, and—in strong coupling regimes—creates polaritonic
states that can redirect chemical reactivity and generate new
spectroscopic routes
\cite{Garcia-Vidaleabd0336,Ruggenthaler2018,
FlickRiveraNarang,doi:10.1021/acs.chemrev.2c00788}.

Recent reviews span quantum optics, chemistry, and materials science
\cite{Ruggenthaler2018,FlickRiveraNarang,doi:10.1021/acs.chemrev.2c00788,Basov2025}.
Ebbesen and collaborators highlight how strong
coupling reshapes chemical landscapes
\cite{doi:10.1021/acs.accounts.6b00295,Garcia-Vidaleabd0336}, while
Rubio
surveys theoretical methods for light-matter hybridization
\cite{Ruggenthaler2018,doi:10.1021/acs.chemrev.2c00788}. DePrince
summarizes advances in
electronic-structure methods for QED chemistry
\cite{foley2023ab,doi:10.1021/acs.jpclett.3c01294},
and Basov provides a materials perspective \cite{Basov2025}.
Together, these reviews
reflect the increasingly interdisciplinary reach of cavity QED
\cite{PhysRevA.90.012508,henry,doi:10.1073/pnas.2110464118,PhysRevLett.115.093001,
PhysRevLett.110.233001,doi:10.1021/acs.jpclett.3c01294,PhysRevLett.125.123602,
PhysRevX.5.041022,10.1063/5.0225932,10.1063/5.0233717,10.1063/5.0230565,10.1063/5.0188471,
D3CP01415K,badanko2022topological,mazza2019superradiant,di2019resolution,galego2015cavity,
shalabney2015coherent,buchholz2019reduced,schafer2019modification,
lacombe2019exact,mandal2020polarized,flick2019light,latini2019cavity,
flick2018strong,garcia2021manipulating,thomas2019tilting,PhysRevB.108.235424,
mordovina2020polaritonic,wang2021light,haugland2021intermolecular,flick2020ab,
doi:10.1073/pnas.1518224112,mandal2020polariton,
tokatly2018conserving,vu2024cavity,rokaj2018light,flick2017atoms}.

In cavity QED, two spatially separated atoms (or molecules) can
interact over distances far beyond their bare dipole–dipole range
because they both couple to the same quantized cavity mode. In the
dispersive regime, virtual photon exchange generates effective
spin–spin or exciton–exciton couplings that can be engineered to
entangle well-separated atom pairs and control their energy-level
structure \cite{PhysRevResearch.7.L012058,PhysRevResearch.6.L012050}.
Experiments and theory now show that this
photon-mediated coupling can hybridize distant molecules into ``optical
bonds,'' where individual emitters share delocalized polaritonic states
via a common microcavity mode \cite{haugland2021intermolecular}.
Real-time simulations of strongly coupled electron–photon dynamics
reveal how energy is redistributed between matter and light degrees of
freedom during such processes, highlighting the role of polariton
formation, dephasing, and dissipation in setting the strength and
coherence of inter-atomic interactions
\cite{castagnola_realistic_2025,welman2025lightmatterentanglementrealtimenuclearelectronic}.

From a more general perspective, off-resonant coupling to the cavity
background field can induce long-range, cavity-mediated interactions
between low-energy excitations even when no real photons are present,
providing a unifying framework for photon-induced modifications of
collective phases and transport
\cite{mazza2019superradiant,feist2015extraordinary}.
Complementary numerical studies of photon-mediated energy transfer
show that the cavity can substantially enhance or suppress transfer
efficiencies between atoms and molecules, depending on spectral
detuning and spatial configuration \cite{Schafer4883,Coles2014}. 
At the many-body
level, these mechanisms give rise to cavity-enhanced energy transport
in extended molecular systems, where excitations delocalize over many
emitters and propagate via polaritonic channels that can span
micrometer scales
\cite{feist2015extraordinary,PhysRevLett.114.196403}. Together, these
works
establish that in appropriately designed cavities, ``light-induced''
interactions are not a small perturbation but a powerful resource for
tailoring how two (or many) atoms talk to each other through the
shared electromagnetic vacuum.

The Pauli--Fierz (PF) Hamiltonian provides a rigorous and
gauge-invariant formulation of the interaction between nonrelativistic
charged matter and the quantized electromagnetic field
\cite{doi:10.1098/rspa.1939.0140,greiner}. It constitutes the
fundamental
starting point for a wide range of theoretical and computational
approaches in cavity quantum electrodynamics (QED)
\cite{ruggenthaler2014quantum,craig1998molecular}, enabling a
consistent treatment
of light--matter coupling across different gauges and levels of
approximation. Owing to the explicit coupling between electronic and
photonic degrees of freedom, the PF Hamiltonian leads to a
substantially enlarged Hilbert space, rendering its exact solution
generally intractable for realistic many-body systems.

In practice, this complexity is addressed by adopting either
wave-function-based or density-based approaches. Wave-function-based
methods explicitly represent the coupled electron--photon state in a
tensor-product Hilbert space and allow for systematic improvability
and high accuracy, albeit at a steep computational cost
\cite{PhysRevX.10.041043,haugland2021intermolecular}. In contrast,
density-based approaches, most
notably quantum electrodynamical density functional theory (QED-DFT),
reformulate the PF problem in terms of reduced variables, yielding
coupled auxiliary equations for matter and photon fields that are
computationally more tractable for extended systems
\cite{ruggenthaler2014quantum,Flick3026}.

Wavefunction-based cavity QED approaches explicitly construct
correlated electron--photon states in a tensor-product Hilbert space
composed of an electronic many-body basis and a photonic Fock basis.
Within this framework, the full PF Hamiltonian is treated
variationally or through systematic many-body expansions, enabling an
accurate and nonperturbative description of strong light--matter
coupling and photon dressing effects.
Representative methods include quantum electrodynamical Hartree--Fock
(QED-HF) \cite{riso_molecular_2022}, coupled-cluster generalizations
such as QED-CC and equation-of-motion QED-CC
\cite{PhysRevX.10.041043,liebenthal_equation--motion_2022}, and
multireference
approaches like QED complete-active-space configuration interaction
(QED-CASCI) \cite{vu_cavity_2024}. In addition, explicitly
correlated stochastic variational methods have been developed to
capture electron--photon correlations beyond standard orbital-based
expansions \cite{PhysRevLett.127.273601,PhysRevA.110.043119}.
These wavefunction-based methods are systematically improvable and can
achieve benchmark-level accuracy for ground and excited polaritonic
states. However, their computational cost grows steeply with the
number of electronic degrees of freedom and photon modes, as the
combined electron--photon Hilbert space scales exponentially. This
limits their applicability to relatively small molecular systems and
few-mode cavities.

Density-based methods, most notably quantum electrodynamical density
functional theory (QED-DFT), reformulate the PF problem in terms of
coupled electron and photon densities rather than explicit many-body
wavefunctions. In this approach, the interacting electron--photon
system is mapped onto auxiliary Kohn--Sham electrons coupled to
quantized or classical Maxwell-like equations for the photon modes
\cite{ruggenthaler2014quantum,tokatly_time_2013}.

The resulting QED-Kohn--Sham equations retain the favorable scaling of
conventional electronic DFT while incorporating light--matter
interactions through effective electron--photon exchange--correlation
functionals \cite{Flick15285}. This framework enables
simulations of realistic molecular and condensed-phase systems in
optical cavities, spanning regimes from weak to strong coupling
\cite{Flick3026}.

Practical implementations often employ reduced cavity descriptions or
effective models, such as single-mode Rabi or Dicke Hamiltonians, to
capture dominant photonic effects at low computational cost
\cite{Jaynes1962ComparisonOQ,PhysRevB.108.235424}. While QED-DFT is
computationally
efficient and scalable, its predictive accuracy depends critically on
the development of reliable exchange--correlation approximations
capable of describing nonclassical electron--photon correlations.

To describe realistic cavity QED systems within the Pauli--Fierz
framework, further physically motivated approximations are typically
introduced in order to render the problem computationally tractable.
One central simplification is the dipole (long-wavelength)
approximation, which exploits the separation of length scales between
the spatial extent of the matter subsystem and the wavelength of the
relevant cavity modes. When the electronic system is localized on a
length scale much smaller than the cavity wavelength, the spatial
dependence of the electromagnetic mode functions can be neglected
across the matter region, and the vector potential may be evaluated at
a fixed reference position. As a result, the light--matter interaction
reduces to a coupling between the total electric dipole moment of the
matter system and the quantized field amplitude, while preserving
gauge invariance at the level of the full Pauli--Fierz Hamiltonian
\cite{greiner,ruggenthaler2014quantum}.

A second widely employed approximation is the truncation of the
electromagnetic field to a finite number of cavity modes, often to a
single resonant mode that dominates the light--matter interaction.
This truncation is physically justified when the cavity features a
well-isolated resonance with a large quality factor and when
off-resonant modes are energetically separated and weakly populated.
However, care must be taken, as a naive mode truncation can violate
gauge invariance and lead to unphysical results, such as spurious
ground-state instabilities or incorrect light--matter decoupling
limits. A consistent single-mode description therefore requires
retaining all gauge-related terms generated by the truncation,
including diamagnetic contributions and self-polarization terms,
ensuring that physical observables remain gauge invariant
\cite{ruggenthaler2014quantum,Flick3026,Taylor:22}. Under these
conditions, the single-mode Pauli--Fierz Hamiltonian provides a
controlled and widely used effective model for realistic cavity QED
systems.

Importantly, no physical cavity is strictly single mode;
rather, this approximation is justified when one cavity mode dominates
the dynamics over all relevant energy and time scales
\cite{Garcia-Vidaleabd0336,pico1,pico2}. Such conditions are realized
in a 
variety of experimental platforms.
In Fabry--P\'erot microcavities, a small mirror separation leads to a
large free spectral range, allowing a single standing-wave mode to be
spectrally isolated from neighboring longitudinal and transverse
modes. Photonic crystal nanocavities provide an even stronger form of
modal confinement, where a localized defect mode appears inside a
photonic bandgap, yielding ultra-small mode volumes and strong
light--matter coupling \cite{Luo2024}. In the microwave domain,
circuit QED architectures achieve exceptionally clean single-mode
behavior through large mode spacings and low dissipation
\cite{Cubaynes2019,Blien2020}. Plasmonic nanocavities, while
intrinsically lossy,
can also be described in terms of an effective single dominant
resonance when dissipation and non-Hermitian effects are properly
accounted for \cite{pico1,pico2}.

In this work, we present a real-time, first-principles investigation
of light-mediated interactions between spatially separated molecules
using time-dependent quantum electrodynamical density functional
theory (QED-TDDFT) in the velocity gauge. Our approach combines a
Pauli–Fierz Kohn–Sham formulation with a tensor-product representation
of electronic real-space orbitals and photonic Fock states, enabling
explicit simulation of coupled electron–photon dynamics beyond
perturbation theory. The electromagnetic field is treated within a
consistent single-mode
approximation, retaining both paramagnetic and diamagnetic couplings
to preserve gauge invariance \cite{10.1063/5.0123909}.

We use a delta-kick excitation to selectively excite only one molecule
in
a system of two widely separated molecules, followed by real-time
propagation of the coupled electron–photon system. The delta kick
represents an ultrashort laser pulse that instantaneously imparts
momentum to the electrons of the targeted molecule while leaving the
second molecule completely unperturbed at the initial time. This
protocol allows us to unambiguously isolate and identify the mechanism
by which excitation can be transferred between distant molecules.

Our results demonstrate a striking contrast between vacuum and cavity
environments. In the absence of light–matter coupling, excitation
remains strictly localized: the molecule subjected to the delta kick
exhibits dipole oscillations, while the second molecule—being far
beyond the range of direct intermolecular interactions—remains
entirely unexcited. In sharp contrast, when both molecules are coupled
to the same cavity mode, the initially excited molecule generates a
cross-coupling via the shared the photon field, which then exterts a back
action on the second, distant molecule. After a short transient, 
the unexcited molecule begins to
oscillate, eventually synchronizing its dipole motion with that of the
initially perturbed molecule. This behavior provides direct, real-time
evidence of photon-mediated excitation transfer between distant
molecular subsystems \cite{Schafer4883,Coles2014,feist2015extraordinary}.

By analyzing the time evolution of molecular dipoles, photon
occupation numbers, photon coordinates, and charge-density
rearrangements, we show that the cavity mode acts as a dynamical
communication channel that converts a strictly local excitation into a
collective response. Importantly, this effect does not rely on
near-field Coulomb interactions, molecular overlap, or classical
radiation fields; it arises solely from the coherent coupling of both
molecules to the quantized cavity mode. The phenomenon persists across
different molecular species (formaldehyde, HF, CO, and mixed dimers),
and its strength and symmetry depend sensitively on molecular
orientation and field polarization
\cite{haugland2021intermolecular,liebenthal_orientation_2024}.

Overall, this work establishes a clear and physically transparent
picture of how ultrashort optical excitation of a single molecule can
trigger coherent dynamics in a distant, initially inactive molecule
via quantized light. Beyond its fundamental significance, this
mechanism opens new possibilities for cavity-controlled energy
transport, molecular synchronization, and nonlocal manipulation of
molecular excitations on ultrafast timescales
\cite{Garcia-Vidaleabd0336,doi:10.1021/acs.chemrev.2c00788}.

\section{Formalism}
The Pauli--Fierz  Hamiltonian provides a gauge-invariant description of
nonrelativistic charged matter interacting with quantized electromagnetic
fields. In the velocity (Coulomb) gauge, the separation between
instantaneous Coulomb interactions and dynamical photon-mediated interactions
is explicit. This makes the velocity gauge particularly well suited for
discussing truncations of the photonic Hilbert space, such as the
single--mode approximation commonly used in cavity quantum electrodynamics.

\subsection{Pauli--Fierz Hamiltonian in Velocity Gauge}
We work in the Coulomb gauge,
$
\nabla \cdot \hat{\mathbf A}(\mathbf r) = 0,
$
and consider nonrelativistic electrons of charge $-e$ and mass $m$ interacting
with the quantized electromagnetic field.

The full Pauli--Fierz Hamiltonian reads
\begin{equation}
\hat H
=
\sum_{i}
\frac{1}{2m}
\left[
\hat{\mathbf p}_i
+ e \hat{\mathbf A}(\hat{\mathbf r}_i)
\right]^2
+
\hat V_{\mathrm{ext}}
+
\hat V_{\mathrm{Coul}}^{\parallel}
+
\hat H_{\mathrm{EM}},
\label{eq:PF_full}
\end{equation}
where
$\hat{\mathbf p}_i = -i\hbar\nabla_i$ is the canonical momentum,
$\hat V_{\mathrm{ext}}$ includes nuclei--electron interactions and any
external scalar potentials.
The instantaneous longitudinal Coulomb interaction,
\begin{equation}
\hat V_{\mathrm{Coul}}^{\parallel}
=
\frac{1}{2}
\int d\mathbf r \, d\mathbf r'
\frac{\hat \rho(\mathbf r)\hat \rho(\mathbf r')}{4\pi\varepsilon_0 |\mathbf r-\mathbf r'|},
\end{equation}
with the electron charge density operator
\begin{equation}
\hat \rho(\mathbf r) = -e \sum_i \delta(\mathbf r-\hat{\mathbf r}_i).
\end{equation}
This term is \emph{independent of the transverse vector potential} and is fixed
by Gauss's law. It must not be altered when truncating the photon
field \cite{hamil}. 

The transverse vector potential is expanded in cavity modes:
\begin{equation}
\hat{\mathbf A}(\mathbf r)
=
\sum_{\alpha}
\sqrt{\frac{\hbar}{2\varepsilon_0\omega_\alpha}}
\,
\mathbf f_\alpha(\mathbf r)
\left(
\hat a_\alpha + \hat a_\alpha^\dagger
\right),
\label{eq:A_mode}
\end{equation}
where $\mathbf f_\alpha(\mathbf r)$ are real transverse mode functions
satisfying the cavity boundary conditions and normalization
(e.g.\ $\int_V d\mathbf r\,|\mathbf f_\alpha(\mathbf r)|^2=1$).
The Hamiltonian of the transverse photon field is defined as
\begin{equation}
\hat H_{\mathrm{EM}}
=
\sum_{\alpha}
\hbar\omega_\alpha
\left(
\hat a_\alpha^\dagger \hat a_\alpha + \frac{1}{2}
\right).
\end{equation}

\subsection{Single--Mode Truncation in Velocity Gauge}

A consistent single--mode approximation is obtained by \emph{projecting the
transverse vector potential onto one cavity mode} $\alpha=c$:
\begin{equation}
\hat{\mathbf A}(\mathbf r)
\;\longrightarrow\;
\hat{\mathbf A}_c(\mathbf r)
=
\sqrt{\frac{\hbar}{2\varepsilon_0\omega_c}}
\,
\mathbf f_c(\mathbf r)
\left(
\hat a + \hat a^\dagger
\right),
\label{eq:A_single}
\end{equation}
and simultaneously truncating the photon Hamiltonian to
\begin{equation}
\hat H_{\mathrm{EM}}
\;\longrightarrow\;
\hbar\omega_c
\left(
\hat a^\dagger \hat a + \frac{1}{2}
\right).
\end{equation}
No modification is made to the longitudinal Coulomb interaction or to the KS
scalar potential $V_{\mathrm{KS}}$.

\subsection{Single--Mode PF--KS Hamiltonian}

The single-particle KS Hamiltonian in the single--mode velocity gauge becomes
\begin{equation}
\hat H_{\mathrm{KS}}
=
\frac{1}{2m}
\left[
- i\hbar \nabla
+ e \hat{\mathbf A}_c(\mathbf r)
\right]^2
+
V_{\mathrm{KS}}(\mathbf r,t).
\label{eq:KS_single_mode}
\end{equation}
Expanding the kinetic term yields
\begin{equation}
\hat H_{\mathrm{KS}}
=
\frac{\hat{\mathbf p}^2}{2m}
+
\frac{e}{m}
\hat{\mathbf p}\cdot\hat{\mathbf A}_c(\mathbf r)
+
\frac{e^2}{2m}
\hat{\mathbf A}_c^2(\mathbf r)
+
V_{\mathrm{KS}}(\mathbf r,t).
\label{eq:KS_single_mode_expanded}
\end{equation}
Both the paramagnetic $\hat{\mathbf p}\cdot\hat{\mathbf A}_c$ and diamagnetic
$\hat{\mathbf A}_c^2$ terms must be retained.

\subsection{Dipole (Long--Wavelength) Approximation}

Assume the electronic system is localized near a point $\mathbf r_0$ (e.g.\ the
molecular center of mass) within a region of characteristic size $\ell$.
If the cavity mode varies on a length scale $L$ such that
\begin{equation}
\ell \ll L,
\end{equation}
then within the matter region the mode profile can be approximated as constant:
\begin{equation}
\mathbf f_c(\mathbf r)\approx \mathbf f_c(\mathbf r_0)\equiv \mathbf f_c^0.
\label{eq:dipole_f}
\end{equation}
This is the \emph{dipole} or \emph{long-wavelength} approximation.

In this approximation the vector potential becomes spatially uniform over the
matter:
\begin{equation}
\hat{\mathbf A}_c(\mathbf r)
\approx
\hat{\mathbf A}_c(\mathbf r_0)
=
\bm{\mathcal A}_c
\left(
\hat a_c+\hat a^\dagger_c
\right),
\qquad
\bm{\mathcal A}_c
\equiv
\sqrt{\frac{\hbar}{2\varepsilon_0\omega_c}}\,\mathbf f_c^0.
\label{eq:A_dipole}
\end{equation}
Often one writes
$
\mathbf f_c^0 = \frac{\bm{\varepsilon}}{\sqrt{V_{\mathrm{eff}}}},
$
where $\bm{\varepsilon}$ is a (unit) polarization vector and $V_{\mathrm{eff}}$
is an effective mode volume at $\mathbf r_0$. 
Introducing the dimensionless photon coordinate
$\hat q_c = (\hat a_c+\hat a_c^\dagger)/\sqrt{2}$, the
vector
potential becomes spatially uniform,
\begin{equation}
\hat{\mathbf A}_c
= \
\sqrt{\frac{\hbar}{\omega_c}}\,
\boldsymbol{\lambda}_c \,\hat q_c,
\qquad
\boldsymbol{\lambda}_c
= \frac{\boldsymbol{\varepsilon}_c}{\sqrt{\varepsilon_0 V}},
\label{eq:A_long_wavelength}
\end{equation}

\subsection{Dipole-Approximate Single--Mode PF--KS Hamiltonian}

Inserting Eq.~\eqref{eq:A_dipole} into Eq.~\eqref{eq:KS_single_mode} yields the
dipole-approximate PF--KS single-particle Hamiltonian
\begin{equation}
\hat H_{\mathrm{KS}}
=
\frac{1}{2m}
\left[
- i\hbar \nabla
+ e \bm{\mathcal A}_c(\hat a_c+\hat a_c^\dagger)
\right]^2
+
V_{\mathrm{KS}}(\mathbf r,t).
\label{eq:KS_single_mode_dip}
\end{equation}
The KS potential is decomposed as
\begin{equation}
V_{\mathrm{KS}}(\mathbf{r}) =
V_{\mathrm{H}}[\rho](\mathbf{r}) +
V_{\mathrm{XC}}[\rho](\mathbf{r}) + V_{\mathrm{ion}}(\mathbf{r}),
\end{equation}
with $V_{\mathrm{H}}$ the Hartree term,
$V_{\mathrm{XC}}$ a GGA exchange–correlation
functional (e.g., Perdew–Burke–Ernzerhof)
\cite{Perdew1996}, and $V_{\mathrm{ion}}$ the external
ionic potential.

\subsection{Tensor-Product KS Ansatz}
We represent the KS orbitals on a tensor-product of
real space and photon Fock space. For electronic
orbital index $m$ and photon number $n$,
\begin{equation}
\Phi_{mn}(\mathbf{r}) = \phi_{mn}(\mathbf{r})
\otimes |n\rangle, \quad n = 0, \dots, N_F,
\end{equation}
where $N_F$ is the Fock truncation (often $N_F=1$
for the vacuum plus single-photon sector). In
practice, the computational domain of size $N_x
\times N_y \times N_z \times N_F$, with
$(N_x,N_y,N_z)$ the real-space grid.

Because Fock states are orthonormal, overlaps factorize:
\begin{equation}
\left( \Phi_{mn} \middle| \Phi_{m'n'} \right) =
\left\langle \phi_{mn} \middle| \phi_{m'n}
\right\rangle \delta_{nn'},
\end{equation}
where the real-space inner product is
\begin{equation}
\left\langle \phi_{mn} \middle| \phi_{m'n}
\right\rangle = \sum_{i,j,k}
\phi_{mn}(x_i,y_j,z_k)\,\phi_{m'n}(x_i,y_j,z_k).
\end{equation}
We orthogonalize the real-space components
within each photon sector (e.g., Gram–Schmidt)
and normalize such that
\begin{equation}
\sum_{n=0}^{N_F}
\left\|\hat{\phi}_{mn}\right\|^2 = 1 \quad
\forall\, m.
\end{equation}

The expansion of the kinetic term in \eqref{eq:KS_single_mode_dip}
contains the
paramagnetic coupling 
\begin{equation}
\frac{e}{m}\hat{\mathbf p}\!\cdot\!\hat{\mathbf A}=
\frac{e}{m}
\sqrt{\frac{\hbar}{\omega_c}} \hat{\mathbf p}\!\cdot\!\boldsymbol{\lambda}_c\hat{q}_c,
\end{equation}
and the diamagnetic (seagull) term 
\begin{equation}
\frac{e^2}{2m}\hat{\mathbf A}^2=
\frac{e^2}{2m}
\frac{\hbar}{\omega_c}
\boldsymbol{\lambda}{_c}^2\hat{q}{_c}^2.
\end{equation}
The paramagnetic interaction links photon states that differ by one
quantum number ($\Delta {n}=\pm 1$)
, while the diamagnetic interaction connects photon states with
quantum number changes of $\Delta n=0,\pm 2$.
In the special case where the diamagnetic term couples photon states
with identical quantum numbers ($\Delta n=0$), it behaves analogously
to the dipole self-interaction (DSI)
found in the length-gauge formulation.

\subsection{Physical quantities}

The full KS orbital is
\begin{equation}
|\Phi_m (t) \rangle
= \sum_{n=0}^{N_F} \phi_{mn}(\mathbf{r,t})
\otimes |n\rangle .
\end{equation}

We define the electronic dipole moment operator of the system as
\begin{equation}
\hat{\mathbf d} = -e\,\hat{\mathbf r},
\end{equation}
where $e$ is the elementary charge and $\hat{\mathbf r}$ is the
electronic
position operator. The total dipole moment (summed over occupied orbitals) is then
\begin{equation}
{\mathbf d}(t) = 
 \sum_{m \in \mathrm{occ}} \langle \Phi_m(t) | \hat{\mathbf
d} | \Phi_m(t) \rangle .
\label{di}
\end{equation}

We define the weight of the $n$-photon sector (summed over
orbitals) as
\begin{equation}
P_n(t) = \sum_m \int d\mathbf{r}\,
\bigl|\phi_{mn}(\mathbf{r,t})\bigr|^2 .
\end{equation}

For a single photon mode with frequency $\omega$ (we drop $c$ in the
following)the
photon coordinate operator is
\begin{equation}
\hat{q} = \sqrt{\frac{\hbar}{2\omega}}\,(\hat a + \hat
a^\dagger) .
\end{equation}
The expectation value of the photon coordinate for the
KS orbital
$|\Phi_m\rangle$ is therefore
\begin{eqnarray}
q_m(t)
&=& \langle \Phi_m(t) | \hat{q} | \Phi_m(t) \rangle \\
&=& \sqrt{\frac{\hbar}{2\omega}}
\sum_{n=0}^{N_F-1} \sqrt{n+1}\left(
\langle \phi_{mn}\vert\phi_{mn+1}\rangle+
\langle \phi_{mn+1}\vert\phi_{mn}\rangle\right)
\nonumber
\end{eqnarray}

\noindent The total photon-coordinate
expectation value is obtained by summing
over orbitals, 
\begin{equation}
q(t) = \sum_m q_m(t) .
\end{equation}
The photon coordinate depends
on the coherence between adjacent
photon-number sectors $\vert n\rangle$ and $\vert n+1\rangle $, not on the
photon-number probabilities $\vert \phi_{mn}\vert^2$ alone.

We can also define the energy of system A
\begin{equation}
E_A(t) = \sum_{m\in A} \langle \Phi_m(t)\vert \hat{H}\vert
\Phi_m(t)\rangle, 
\end{equation}
where the summation includes orbitals belong to system A. This will be
used to characterize energy transfer between two molecules. The dipole
of a given subsystem can be defined similarly to \eqref{di}
restricting the summations to the orbitals belonging to A.

\subsection{Time propagation}
The real-time evolution of each Kohn--Sham (KS) orbital $\Phi_m(t)$ is
governed
by the time-dependent Schr\"odinger equation
\begin{equation}
i\hbar \frac{\partial}{\partial t} \ket{\Phi_m(t)}
=
\hat{H}_{KS}(t)\ket{\Phi_m(t)}
\label{eq:TDKS}
\end{equation}

The initial KS orbitals are obtained by minimizing the ground-state
Kohn--Sham energy functional defined on the coupled electron--photon
Hilbert space. This minimization is carried out using iterative
gradient-based schemes, where the Pauli--Fierz--augmented Hamiltonian is
assembled in a tensor-product representation of electronic real-space
orbitals and photonic Fock states.

To access excitation spectra, the
system is perturbed at the initial time by an impulsive (delta-kick)
perturbation. In the velocity gauge, the excitation is implemented via a spatially
uniform, time-dependent vector potential, 
\begin{equation} \mathbf
A(t) = - \kappa \hat{\mathbf e} \Theta(t), 
\end{equation} 
where
$\kappa$ controls the strength of the perturbation, $\hat{\mathbf e}$
is the polarization direction, and $\Theta(t)$ is the Heaviside step
function. The associated electric field is given by $\mathbf E(t) =
-\partial_t \mathbf A(t) = \kappa \hat{\mathbf e} \delta(t)$.

In the PF Hamiltonian formulated in the velocity gauge, the
coupling to the vector potential enters through the minimal-coupling
substitution. The action of the delta kick is therefore realized by an
instantaneous boost of the electronic momenta. At the level of the KS
orbitals, this corresponds to the transformation 
\begin{equation}
\Phi_m(0^+) = e^{- i \kappa \hat{\mathbf e}\cdot \mathbf r} 
\Phi_m(0^-), \label{eq:delta_kick_vg} 
\end{equation}
which is
equivalent, up to a gauge transformation, to the length-gauge
formulation. For sufficiently small $\kappa$, this procedure excites
all dipole-allowed modes simultaneously.

In cavity QED simulations, the velocity-gauge delta kick provides a
numerically convenient and gauge-consistent way to initiate coupled
electron--photon dynamics, particularly when paramagnetic and
diamagnetic light--matter coupling terms are treated explicitly.

The real-time propagation of the Kohn--Sham orbitals is performed using
an explicit fourth-order Taylor expansion of the time-evolution
operator,
a method that has proven robust and efficient for real-space
time-dependent density-functional theory and its extensions
\cite{YabanaBertsch1996,Marques2003}. The accuracy and stability of this
approach rely on a consistent choice of the time step and spatial grid
spacing, as discussed in detail in the context of computational
nanoscience simulations \cite{VargaComputationalNanoscience}.

Explicitly, the KS orbitals are advanced according to
\begin{equation}
\ket{\Phi_m(t + \Delta t)} =
\sum_{k=0}^{4} \frac{1}{k!}
\left(
\frac{i\Delta t}{\hbar} \hat{H}(t)
\right)^k
\ket{\Phi_m(t)}
\label{eq:Taylor4}
\end{equation}

    The numerical accuracy and stability of the propagation require a
    consistent choice of the time step $\Delta t$ and the spatial grid
    spacing $\Delta x$. On a real-space grid, the highest kinetic-energy
    components that can be represented are determined by the grid
    spacing,
    \begin{equation}
    E_{\text{max}} \sim \frac{\hbar^2}{2m(\Delta x)^2}.
    \label{eq:Emax}
    \end{equation}
    To accurately resolve these components, the time step must satisfy
    \begin{equation}
    \Delta t \ll \frac{\hbar}{E_{\text{max}}}
    \sim \frac{2m}{\hbar}(\Delta x)^2.
    \label{eq:dt_condition}
    \end{equation}
    When this condition is fulfilled, the fourth-order Taylor propagator
    provides stable and accurate real-time dynamics of the KS orbitals
    in
    coupled electron--photon systems.

\section{Results}
\subsection{Nitrobenzene in cavity coupled to light}

\begin{figure}[h]
\centering
\includegraphics[width=0.5\textwidth]{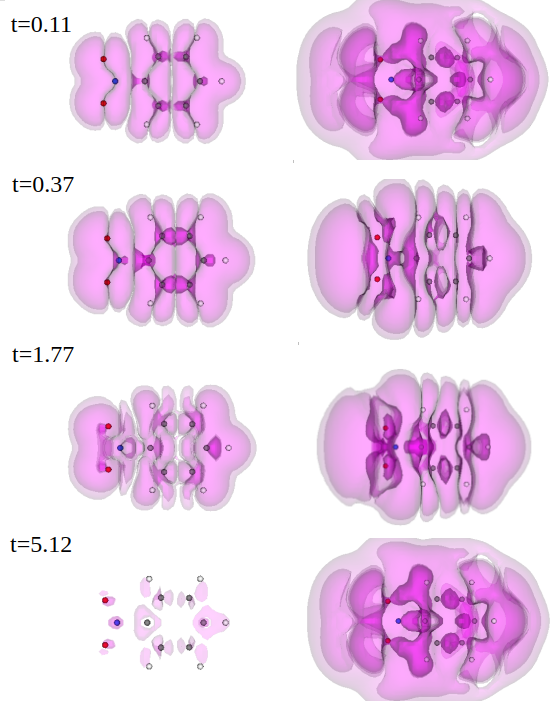}
\caption{Temporal dynamics of nitrobenzene's dipole moment. Solid
line: cavity-coupled molecule; dashed line: uncoupled molecule in
vacuum (magnified by 100 for clarity). $\lambda=0.005$ a.u. and $\omega=$
0.289 a.u. is used in the calculations.}
\label{fig:nitro_dp}
\end{figure}

\begin{figure}[h]
\centering
\includegraphics[width=0.5\textwidth]{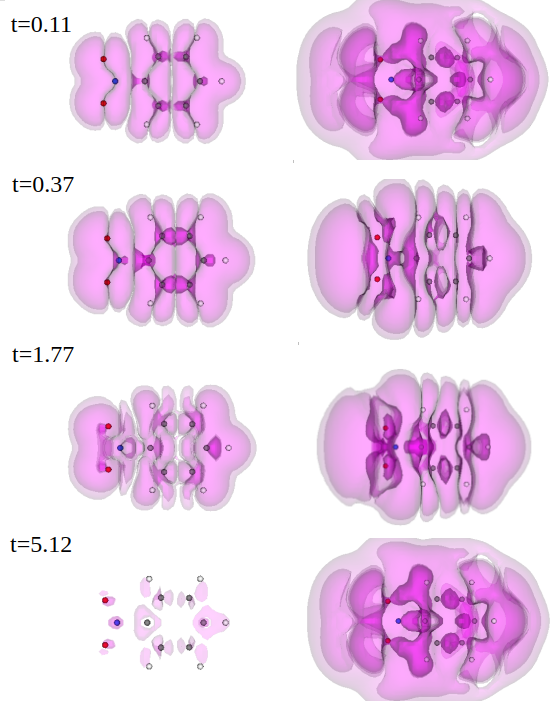}
\caption{
Snapshots of the charge density difference between the excited state
and the ground state. Left panel: molecule in vacuum; right panel:
molecule in an optical cavity. All parameters are identical to those
used in Fig.~\ref{fig:nitro_dp}. The three shades of magenta
represents values of -0.00002,-0.00001,0.00002 electrons/$\Delta x^3$ 
ranging from the lightest to the darkest.}
\label{fig:nitro}
\end{figure}

We computed the ground state of a single Nitrobenzene molecule both in
vacuum and within an optical cavity. Following convergence of the
ground-state calculation, the system was perturbed by a delta-kick
pulse, and the molecular orbitals were propagated in time.
Fig.~\ref{fig:nitro_dp} displays the time evolution of the dipole moment for
Nitrobenzene, plotted as $d = d(t) - d_0$, where $d_0$ represents the
ground-state dipole moment. Two scenarios are compared: the molecule
inside an optical cavity (with light-matter coupling) and the molecule
in vacuum (without coupling). The results reveal that cavity coupling
substantially amplifies the dipole response.
To further illustrate the contrast between these two environments, we
analyzed snapshots of the electron density difference, obtained by
subtracting the ground-state density from the time-dependent
excited-state density. Fig.~\ref{fig:nitro} demonstrates that 
density oscillations
differ markedly between the cavity and vacuum cases. In the cavity,
the density oscillations are more pronounced and localized to distinct
regions of the molecule. While the oscillations decay over time in
vacuum, they persist with significant amplitude in the cavity
environment.


\begin{figure}[h]
\centering
\includegraphics[width=0.5\textwidth]{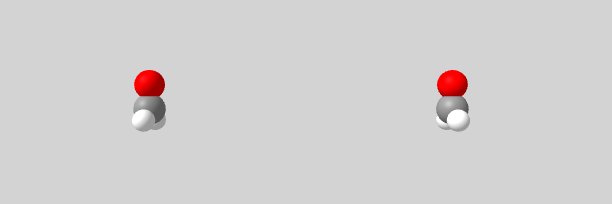}
\caption{
Formaldehyde dimer configuration: cavity frequency $\omega$ = 0.289,
light-matter coupling strength $\lambda$ = 0.005. The light polarization
vector is aligned along the O-C bond direction of each molecule, with
the O-C axes positioned perpendicular to the z-axis that connects the
two carbon centers.
}
\label{fig:form}
\end{figure}

\begin{figure}[htbp]
\centering

\begin{subfigure}[b]{0.45\textwidth}
\centering
\includegraphics[width=\textwidth]{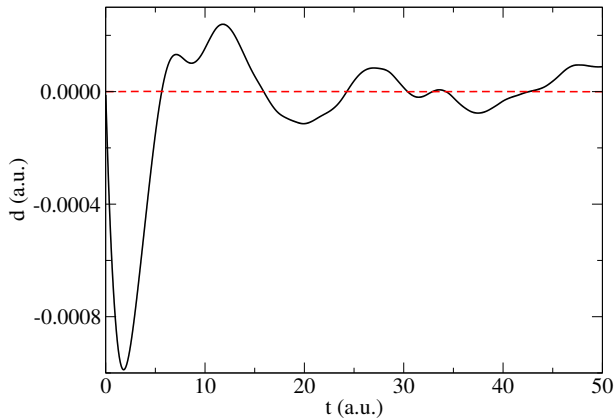}
\caption{
Time evolution of dipole moments following delta kick excitation of
the left molecule only, with no light-mediated coupling between
molecules: left molecule (solid black), right molecule (dashed red).
}
\label{fig:f1}
\end{subfigure}
\hfill
%
\begin{subfigure}[b]{0.45\textwidth}

\centering
\includegraphics[width=\textwidth]{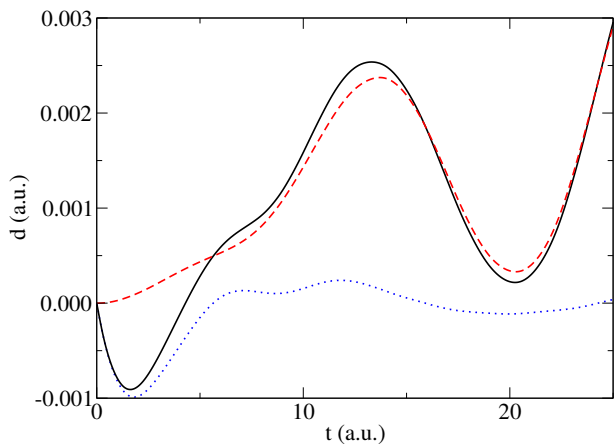}
\caption{
Dipole moment dynamics with light-mediated coupling after selective
delta kick excitation of the left molecule: left molecule (solid
black), right molecule (dashed red). The blue dotted line shows the
uncoupled evolution of the left molecule for comparison, identical to
the black curve in the previous figure.
}
\label{fig:f2}
\end{subfigure}
\hfill	        
\caption{Temporal evolution of dipole moment changes relative to ground
state values for the formaldehyde dimer.
}
\label{fig:formaldehyde1}
\end{figure}

\begin{figure}[htbp]
\centering
\includegraphics[width=0.5\textwidth]{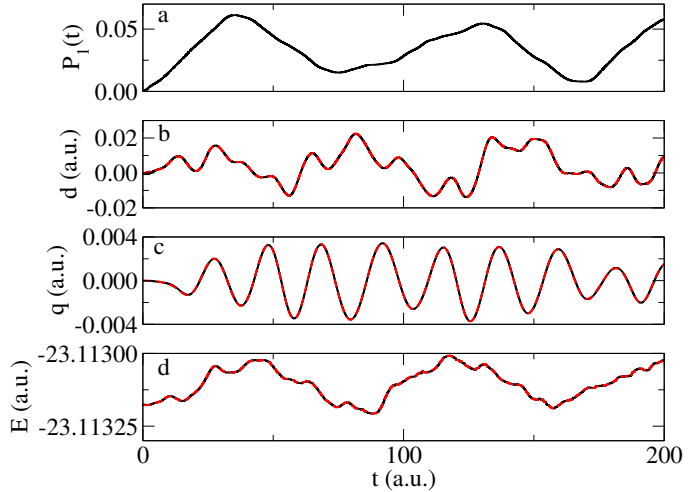}
\caption{Temporal evolution of (a) photon occupation number P$_1$ of the
$\vert 1 \langle$ space, (b)
dipole moment, (c) photon coordinate and (d) energies of 
the formaldehyde dimer. The solid black line shows the dynamics of the 
left, kicked molecule, the dashed red line represents the right
initially unperturbed system.}
\label{fig:formaldehyde}
\end{figure}

\subsection{Two formaldehyde molecules coupled by light}
Next, we examine the light-matter interaction between two spatially
separated molecules. The system consists of two formaldehyde 
molecules separated by
a large distance (30 a.u.)
(see Fig. \ref{fig:form}) such that their direct interaction (without light
coupling) is negligible. Formaldehyde has a substantial dipole moment
of 2.33 Debye, enabling strong light coupling. The left molecule is
selectively excited with a delta kick, and the system dynamics are
time-propagated.
In the first case, where molecules are not coupled to light, Fig.
\ref{fig:f1}
demonstrates that only the excited left molecule's dipole moment
evolves, while the right molecule's dipole remains unchanged at zero.
Fig \ref{fig:f2} compares the dipole dynamics when molecules are
light-coupled, with only the left molecule receiving delta kick
excitation. Initially, the left molecule's dipole begins oscillating
due to the excitation, while the unexcited right molecule's dipole
remains near zero for the first 2-3 atomic units of time. After
approximately 5 a.u., the right molecule's dipole also begins to
oscillate, closely following the left molecule's behavior. For
comparison, this figure includes the uncoupled left molecule evolution
(blue dotted line). Initially, the coupled and uncoupled systems
behave similarly, but after ~5 a.u., light-mediated coupling causes
the systems to diverge.

Fig. \ref{fig:formaldehyde} extends the coupled dynamics over a longer time
period, revealing that after the initial few atomic units, the two
molecular dipoles evolve nearly identically, demonstrating strong
synchronization through light-mediated coupling. The figure
illustrates the temporal evolution of the photon occupation number
(Fig.\ref{fig:formaldehyde}a),
dipole moment(Fig.\ref{fig:formaldehyde}b), photon coordinate
(Fig.\ref{fig:formaldehyde}c), and energies
(Fig.\ref{fig:formaldehyde}d) as the formaldehyde
dimer molecules move in a concerted, identical manner. 
An interesting observation is that the photon occupation number and
photon coordinate, though related, oscillate with very different
periods. The dipole moment—a real-space variable—also oscillates, but
its behavior differs from both quantities. The energy, however, shows
an oscillation pattern more similar to that of the photon occupation
number. This can be easily understood: the photon occupation
corresponds to population changes between the zero-photon and
one-photon states, which have different energies. Consequently, the
total energy oscillates synchronously with these occupation changes.

\begin{figure}[h]
\centering
\includegraphics[width=0.5\textwidth]{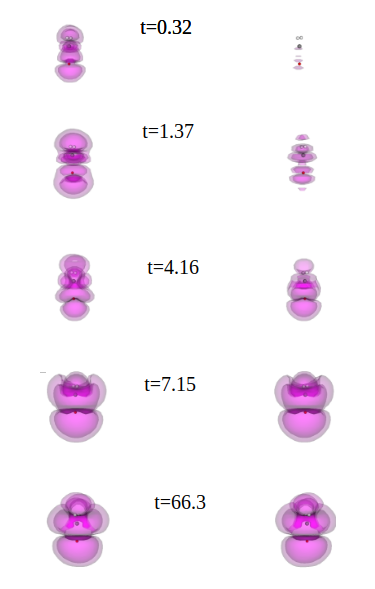}
\caption{
Snapshots of the charge density difference between the excited state
and the ground state for the COH$_2$ molecule. The left molecule is
excited with a delta kick perturbation. $\lambda=0.005$ a.u. and
$\omega=0.289$ a.u. is used in the calculations.
The three shades of magenta
represents values of -0.00005,-0.0002,0.00005 electrons/$\Delta x^3$
ranging from the lightest to the darkest.}
\label{fig:formal}
\end{figure}
Fig. \ref{fig:formal} shows temporal snapshots of charge oscillations for both
molecules. At early times, only the perturbed molecule displays
appreciable oscillations, while charge rearrangement gradually builds
up on the second molecule. Eventually, the charge oscillations
synchronize, with both molecules oscillating identically, as evidenced
by the dipole oscillation (Fig. \ref{fig:formaldehyde}b).

\subsection{Two HF molecules coupled by light}
\begin{figure}[h]
\centering
\includegraphics[width=0.5\textwidth]{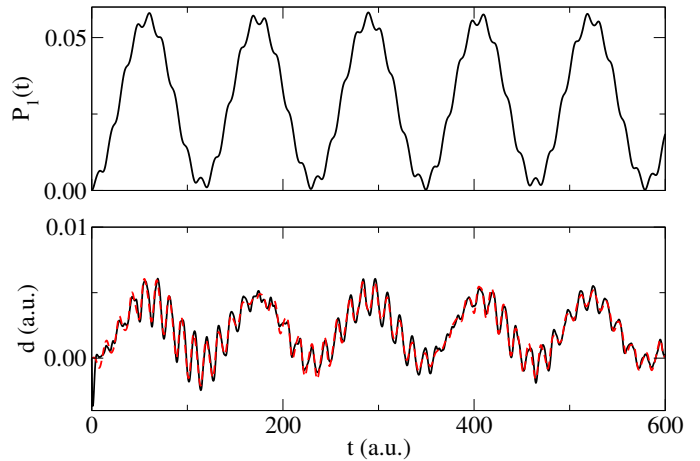}
\caption{Occupation of the $\vert 1 \rangle$ Fock space and the 
change of the dipole moment as the function of time. The solid black
line shows the dipole of the molecule which was perturbed by a delta
kick excitation, the red dashed line shows the dipole of the second
molecule.
$\lambda=0.01$ a.u. and
$\omega=0.467$ a.u. is used in the calculations.}
\label{fig:hfx}
\end{figure}

Next, we present a similar calculation using two spatially separated
HF molecules. 
The molecules are aligned parallel to each other and to
the x-axis, with a separation of 30 a.u. along the perpendicular
z-direction. 
The electromagnetic field is polarized along the x-axis,
parallel to the molecular axes. Fig. \ref{fig:hfx} displays the occupation of
the $\vert 1\rangle$ Fock state and the time evolution of the dipole moment. As
observed in the formaldehyde case, the dipole moments of both
molecules oscillate synchronously after an initial transient period
during which the perturbed molecule induces charge oscillations in the
second molecule. Here, the dipole oscillations exhibit periodic
sinusoidal behavior, and the Fock state occupation follows this
oscillatory pattern. This contrasts sharply with the formaldehyde
case, where no such strong coupling between the dipole and photon
occupation dynamics was observed. 

\begin{figure}[h]
\centering
\includegraphics[width=0.5\textwidth]{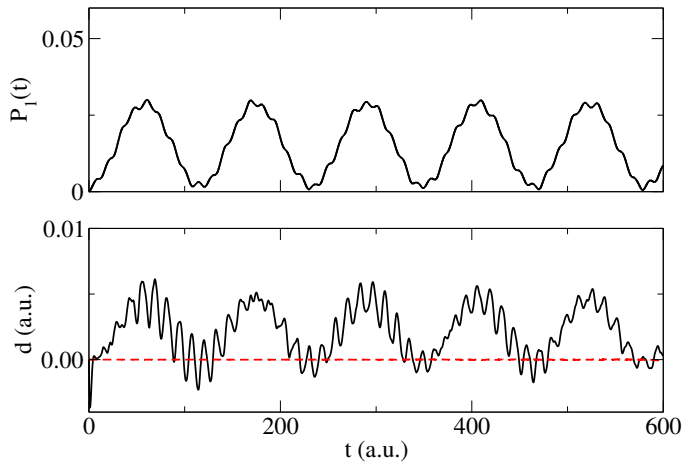}
\caption{Two HF molecules oriented perpendicularly. 
Temporal evolution of (a) photon occupation number, (b)
dipole moment, (c) photon coordinate and (d) energies of 
the formaldehyde dimer. The solid black line shows the dynamics of the 
left, kicked molecule, the dashed red line represents the right
initially unperturbed system. $\lambda=0.01$ a.u. and
$\omega=0.467$ a.u. is used in the calculations.}
\label{fig:hf_ort}
\end{figure}

The coupling between the two molecules is highly sensitive to their
relative orientation and the polarization of the incident light.
Fig. \ref{fig:hf_ort} illustrates the dipole moment evolution 
when the molecule
excited by the delta kick is aligned parallel to the light
polarization—identical to the previously discussed parallel
configuration—while the second unperturbed molecule is oriented along
the z-axis, perpendicular to the first molecule. The dipole moment of
the first molecule exhibits the same temporal behavior as before,
whereas the dipole moment of the second molecule remains nearly
constant over time. Fig. \ref{fig:hf_ort} also displays the 
occupation dynamics of
the $\vert 1\rangle$ Fock state. While the occupation pattern 
resembles that of the
parallel case, its amplitude is reduced by half due to the negligible
contribution from the second molecule.

\begin{figure}[h]
\centering
\includegraphics[width=0.5\textwidth]{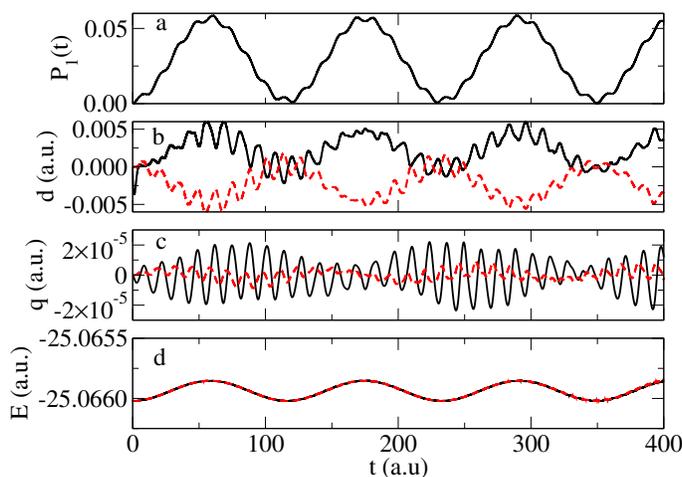}
\caption{Two HF molecules oriented antiparallel.
Temporal evolution of (a) photon occupation number, (b)
dipole moment, (c) photon coordinate and (d) energies of
the formaldehyde dimer. The solid black line shows the dynamics of the
left, kicked molecule, the dashed red line represents the right
initially unperturbed system. $\lambda=0.01$ a.u. and
$\omega=0.467$ a.u. is used in the calculations.}
\label{fig:hf_ap}
\end{figure}

The orientation dependence is further demonstrated in Fig.
\ref{fig:hf_ap}. In
this configuration, the two molecules are aligned antiparallel to each
other, with their molecular axes perpendicular to the field
polarization. The occupation of the $\vert 1\rangle$ Fock state 
remains largely
consistent with the parallel case (Fig.\ref{fig:hf_ap}a); however, the dipole moments now
oscillate synchronously in opposite directions (Fig.\ref{fig:hf_ap}b). 
The photon coordinates also oscillate in the opposite direction, but
with different amplitudes (Fig.\ref{fig:hf_ap}c).
Finally, the energy, which is primarily governed by the photon
occupation number, oscillates in phase with it
(Fig.\ref{fig:hf_ap}d).

\begin{figure}[h]
\centering
\includegraphics[width=0.5\textwidth]{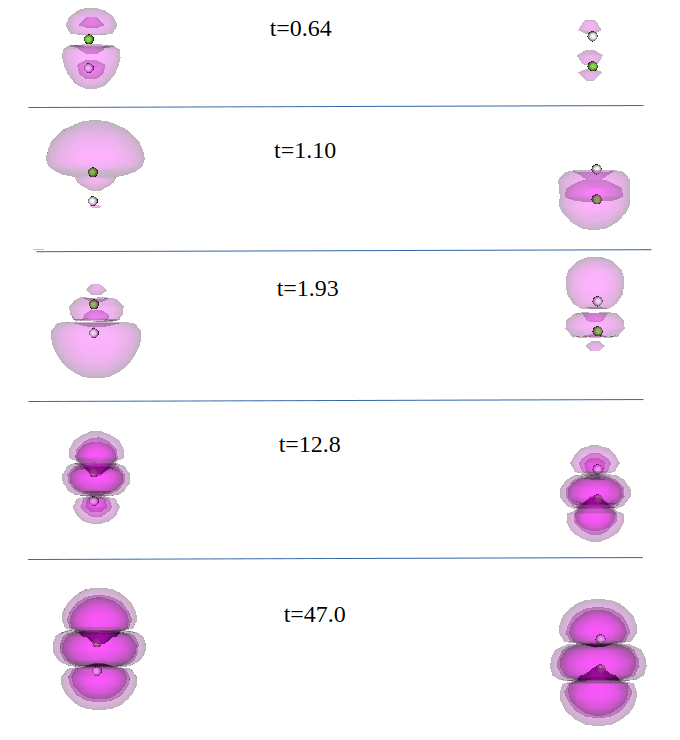}
\caption{
Snapshots of the charge density difference between the excited state
and the ground state for the antiparallel HF molecules. The left molecule is
excited with a delta kick perturbation.
The three shades of magenta
represents values of -0.00003,-0.00001,0.00004 electrons/$\Delta x^3$
ranging from the lightest to the darkest.}
\label{fig:hf}
\end{figure}
Fig. \ref{fig:hf} displays snapshots of the charge density variations 
for the antiparallel-oriented HF molecules. As observed in the formaldehyde
case, excitation initially occurs exclusively in the first molecule,
followed by gradual excitation of the second molecule. Once the second
molecule becomes excited, the charge density oscillations on the two
molecules occur in opposition to one another, mirroring each other's
behavior.

\subsection{Two CO molecules coupled by light}
\begin{figure}[h]
\centering
\includegraphics[width=0.5\textwidth]{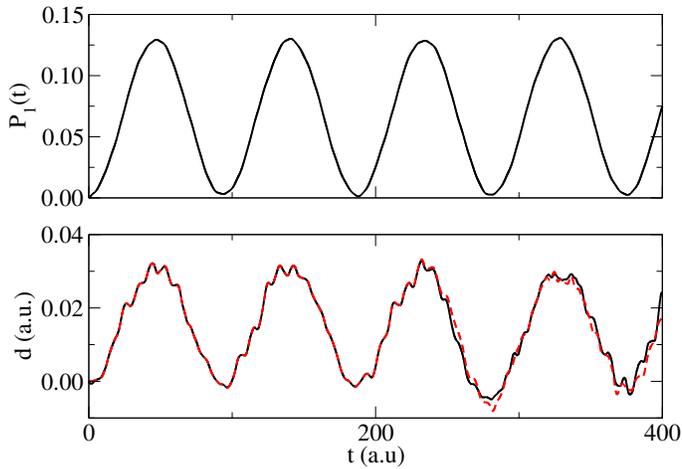}
\caption{Occupation of the $\vert 1 \rangle$ Fock space and the 
change of the dipole moment as the function of time when the CO molecules
oriented parallel. The solid black
line shows the dipole of the molecule which was perturbed by a delta
kick excitation, the red dashed line shows the dipole of the second
molecule.
$\lambda=0.01$ a.u. and
$\omega=0.467$ a.u. is used in the calculations.}
\label{fig:COx}
\end{figure}

\begin{figure}[h]
\centering
\includegraphics[width=0.5\textwidth]{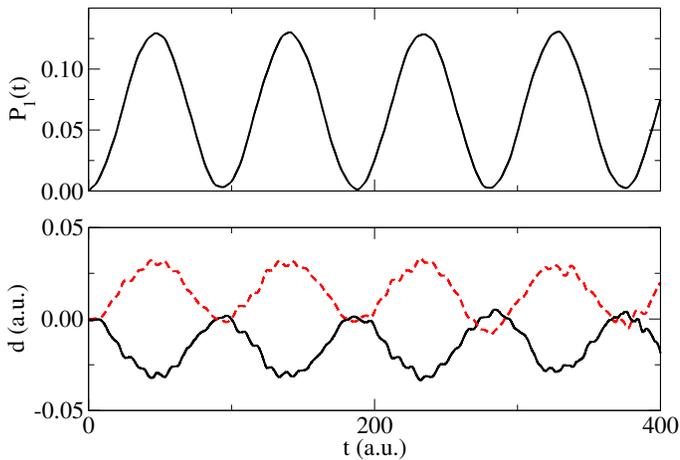}
\caption{Occupation of the $\vert 1 \rangle$ Fock space and the 
change of the dipole moment as the function of time when the CO molecules
oriented antiparallel. The solid black
line shows the dipole of the molecule which was perturbed by a delta
kick excitation, the red dashed line shows the dipole of the second
molecule.
$\lambda=0.01$ a.u. and
$\omega=0.467$ a.u. is used in the calculations.}
\label{fig:COr}
\end{figure}

We conducted a similar investigation to that described in the previous
section, but using CO molecules instead. Fig. \ref{fig:COx} demonstrates that
the molecular dipoles align and oscillate synchronously, mirroring the
behavior observed for parallel HF molecules. However, both the
amplitude and frequency of these oscillations differ compared to the HF
case. The photon occupation in the $\vert 1\rangle$ state is higher for CO
compared to HF, which can be attributed in part to CO having a greater
number of molecular orbitals than HF. Consistent with the HF results,
antiparallel CO molecules exhibit dipole oscillations that are out of
phase with each other (Fig. \ref{fig:COr}). Additionally, the dipole oscillation
of the initially perturbed molecule remains largely unchanged
regardless of whether the molecules are arranged in parallel or
antiparallel configurations. The $\vert 1 \rangle$ photon state 
occupation also shows minimal variation between these two orientations.


\subsection{A CO and a HF molecule coupled by light}

\begin{figure}[h]
\centering
\includegraphics[width=0.5\textwidth]{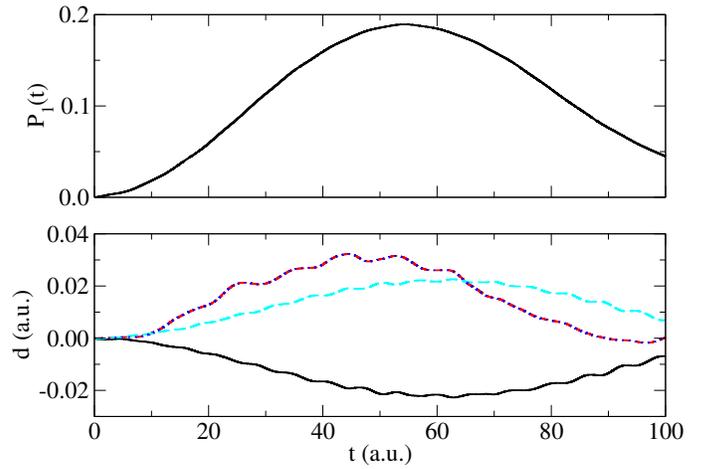}
\caption{Occupation of the $\vert 1 \rangle$ Fock space and the 
change of the dipole moment as the function of time.
The black solid line represents the time-dependent dipole moment
change for the perturbed HF molecule, while the red dashed line
indicates the dipole moment change for the unperturbed CO molecule. When
the H and F atomic positions are reversed, the long dashed cyan line
displays the resulting time-dependent dipole moment change for the
perturbed HF molecule, and the dotted blue line shows the dipole
moment change for the unperturbed CO molecule (which overlaps with the red
dashed line).
$\lambda=0.01$ a.u. and
$\omega=0.467$ a.u. is used in the calculations.}
\label{fig:cohf1}
\end{figure}
\begin{figure}[h]
\centering
\includegraphics[width=0.5\textwidth]{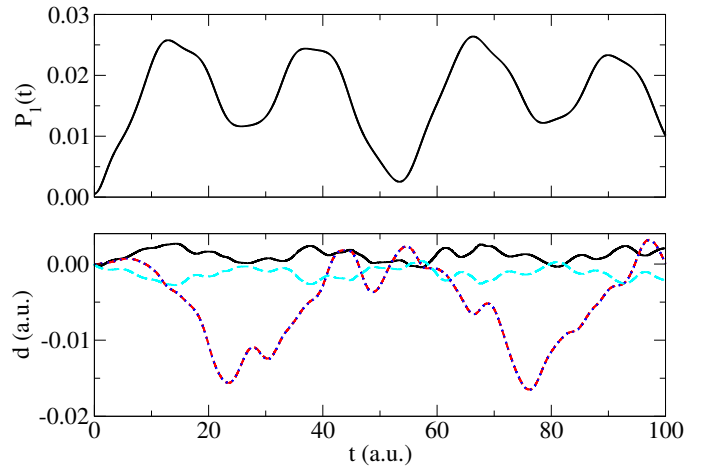}
\caption{Same as Fig. \ref{fig:cohf1} but with
$\omega=0.467$ a.u.}
\label{fig:cohf2}
\end{figure}
In this section, we examine light-mediated coupling between two
distinct molecules. 
The system consists of HF and CO molecules with their molecular axes
aligned parallel to both the x-axis and the light polarization
direction. These molecules are arranged symmetrically about the
z-axis, with each positioned at a distance equal to half its bond
length from the z-axis. A separation of 30 a.u. between the molecules
ensures negligible intermolecular interaction.

A delta kick excitation is applied to an HF
molecule, which subsequently induces excitation in a CO molecule.
Fig. \ref{fig:cohf1} displays the time-dependent changes in dipole moments and the
occupation of the $\vert 1\rangle$ Fock state. Two calculations are presented: in
the first, the F and C atoms are positioned at the top while the H and
O atoms are at the bottom; in the second, the H and F atoms are
exchanged. This exchange leaves the photon occupation unchanged but
symmetrically inverts the dipole oscillation of the HF molecule. The
CO molecule's dipole oscillation remains unaffected in both cases.
Fig. \ref{fig:cohf2} presents an analogous study with a modified cavity frequency
of 0.289 a.u. (compared to 0.467 a.u. used in all previous HF and CO
calculations). The qualitative behavior persists: photon occupation
remains independent of atomic switching, the CO dipole is unchanged,
and the HF dipole undergoes a sign reversal upon switching. However,
the oscillation patterns differ markedly—the photon occupation now
exhibits four oscillations in the same time period compared to the previous single
oscillation, and the dipole moments display increased oscillatory
character.

\section{Summary}

This work presents a real-time investigation of light-mediated
interactions between spatially separated molecules using quantum
electrodynamical time-dependent density functional theory based on the
Pauli–Fierz Hamiltonian in the velocity gauge. We have developed a
first-principles computational framework that combines a Kohn–Sham
formulation with a tensor-product representation of electronic
real-space orbitals and photonic Fock states, enabling explicit
simulation of coupled electron–photon dynamics beyond perturbation
theory. The approach employs a consistent single-mode approximation
within the dipole regime, retaining both paramagnetic and diamagnetic
coupling terms to preserve gauge invariance. Time propagation is
performed using a fourth-order Taylor expansion of the time-evolution
operator, providing a computationally tractable yet rigorous
description while maintaining the favorable scaling of conventional
density functional theory.

Our simulations demonstrate a striking contrast between vacuum and
cavity environments. For Nitrobenzene, cavity coupling substantially
amplifies the dipole response and sustains coherent charge-density
oscillations that decay rapidly in vacuum, reflecting modification of
the molecular energy landscape through polariton formation. For
spatially separated molecular dimers positioned far beyond the range
of direct intermolecular interactions, excitation remains strictly
localized in vacuum, with the distant molecule remaining entirely
unexcited. In sharp contrast, when both molecules couple to the same
cavity mode, the initially excited molecule generates
cross-coupling through the photon field, which acts back on the distant 
molecule. After a transient period, the two molecular dipoles synchronize, oscillating
nearly identically and demonstrating coherent energy exchange mediated
by the shared quantized field.

The strength and symmetry of cavity-mediated coupling depend
sensitively on molecular orientation relative to field polarization.
Parallel-aligned molecules exhibit in-phase dipole oscillations, while
antiparallel configurations produce out-of-phase oscillations with
inverted charge-density dynamics. Perpendicular orientations suppress
coupling almost entirely. The photon occupation number and photon
coordinate exhibit distinct oscillatory behaviors, with the total
energy oscillating synchronously with occupation changes reflecting
population redistribution between photon states. The amplitude,
frequency, and coherence time of cavity-mediated oscillations vary
across different molecular species, reflecting differences in dipole
moments, electronic structure, and resonance conditions. These results
provide direct evidence that the cavity mode functions as a dynamical
communication channel, converting strictly local excitations into
collective responses through coherent coupling to the quantized field
rather than near-field Coulomb interactions or molecular overlap.

The present work establishes a foundation for exploring several
important directions. The tensor-product framework can be scaled to
investigate cavity-mediated energy transport in larger molecular
assemblies, including molecular crystals and light-harvesting
complexes, revealing how excitations delocalize over many emitters via
polaritonic channels. Extending the formalism to multimode
environments will enable more accurate modeling of experimental
platforms and could uncover interference effects and selective
energy-transfer pathways. Incorporating dissipation through cavity
losses and molecular vibrational coupling will be essential for
quantitative comparison with experiments and assessing the robustness
of cavity-mediated phenomena under realistic conditions.
Exploring alternative excitation schemes
beyond delta kicks, such as continuous-wave driving or
phase-controlled pulse sequences, could reveal new regimes of
molecular synchronization and coherent control.


\end{document}